\documentclass[conference]{IEEEtran}
\IEEEoverridecommandlockouts
\usepackage{cite}
\usepackage{amsmath,amssymb,amsfonts}
\usepackage{algorithmic}
\usepackage{physics}
\usepackage{verbatim}
\usepackage{float}
\usepackage{graphicx}
\usepackage{textcomp}
\usepackage{xcolor}
\def\BibTeX{{\rm B\kern-.05em{\sc i\kern-.025em b}\kern-.08em
    T\kern-.1667em\lower.7ex\hbox{E}\kern-.125emX}}
\begin{document}

\title{Noise Dynamics in the Quantum Regime}

\author{
\IEEEauthorblockN{Clovis Farley}
\IEEEauthorblockA{
\textit{Institut Quantique} \\
\textit{Département de physique} \\
\textit{Université de Sherbrooke}\\
Sherbrooke, Canada \\
clovis.farley@usherbrooke.ca}
\and
\IEEEauthorblockN{Edouard Pinsolle}
\IEEEauthorblockA{
\textit{Institut Quantique} \\
\textit{Département de physique} \\
\textit{Université de Sherbrooke}\\
Sherbrooke, Canada \\
edouard.pinsolle@usherbrooke.ca}
\and
\IEEEauthorblockN{Bertrand Reulet}
\IEEEauthorblockA{
\textit{Institut Quantique} \\
\textit{Département de physique} \\
\textit{Université de Sherbrooke}\\
Sherbrooke, Canada \\
bertrand.reulet@usherbrooke.ca
    }
}

\maketitle

\begin{abstract}
A time-dependent bias voltage on a tunnel junction generates a time-dependent modulation of its current fluctuations, and in particular of its variance. This translates into an excitation at frequency  $\tilde f$ generating correlations between current fluctuating at any frequency $f$ and at frequency $\pm\tilde f -f$. We report the measurement of such a correlation in the fully quantum regime, i.e. when both frequencies are much greater than $k_BT/h$ with $T$ the temperature. Such a correlator, usually referred to as the noise susceptibility, is involved in corrections to the measurements of higher-order moments and in the squeezing of noise.
\end{abstract}

\section{Introduction}

The tunnel junction, as one of the simplest coherent conductor, is the perfect platform to study noise dynamics, i.e. the dynamical response of the statistics of current fluctuations to a time-dependent parameter, here the voltage bias. The ac conductance $G(f)$ measures the linear response at frequency $f$ of the average current to a small voltage oscillating at frequency $\tilde f$, with $\tilde f=f$. Going one step further, the noise susceptibility $\chi_{\tilde{f}}(f)$, describes the linear response of the variance of current fluctuations observed at frequency $f$ to an excitation at frequency $\tilde{f}$. This response oscillates at frequency $\tilde f$, not to be confused with the photo-assisted noise which is the dc response of the variance measured at frequency $f$ to the same oscillation, and that is proportional to the square of the ac bias at low bias. The response of the probability distribution should depend on the internal dynamics of transport, but in the case of a tunnel junction, quantum tunneling is almost instantaneous when compared to measurement frequency in the microwave-domain. As a consequence, the response is in phase with the excitation, i.e. the ac conductance and noise susceptibility are real numbers (considering the parasitic capacitance of the junction as external). For a slow excitation, or equivalently a high temperature such that $k_B T \gg h\tilde f$, the distribution follows adiabatically the excitation, so $G(\tilde f)=dI/dV$ with $I$ the (average) dc current and $\chi_{\tilde{f}}(f)=\frac{dS_2(f)}{dV}$ with $S_2(f)$ the noise spectral density of current fluctuations at frequency $f$. $V$ is the voltage across the sample. In contrast, at low temperatures $T \ll h\tilde f/k_B$, quantum effects such as the Pauli exclusion principle and the Heisenberg uncertainty principle affect the electron transport by adding quantum correlations \cite{BlanterButtiker,Thibault2015}, thus modifying $\chi_{\tilde{f}}(f)$.\\

The noise susceptibility has been theoretically analyzed \cite{Gabelli2008_theo}, and shown to be given by the correlator $ \ev{ I(f)I(\tilde f-f) }$ properly symmetrized. The analytical result is given by

\footnotesize
\begin{equation}
    \chi_{ \tilde f }(f) = \frac{e}{2h \tilde f} \qty( S_2^{0}(f_+) - S_2^{0}(f_-) + S_2^{0}(f_- - \tilde f) - S_2^{0}(f_+ - \tilde f) )
    \label{chi}
\end{equation}
\normalsize
\noindent with $f_{\pm} = f \pm \frac{eV}{h}$ and $V$ the dc bias voltage applied to the sample. $S_2^0(f)$ is the equilibrium current noise spectral density, as given by the fluctuation-dissipation theorem \cite{Herbert1951}:
\begin{equation}
    S_2^0(f) = G hf \coth \qty( \frac{hf}{2k_B T} )
\end{equation}
The noise susceptibility has already been measured for $\tilde f \sim f$ by measuring correlations between high frequency fluctuations and low frequency ones, i.e. $ \ev{ I(f)I(0) }$ \cite{Gabelli2008_exp}. Here we report the measurement of $\chi_{\tilde{f}}(f)$ in a fully quantum regime, when all frequencies involved ($f$, $\tilde f$, $\tilde f+f$ and $\tilde f-f$) are much greater than $k_B T/h$. More precisely, we report the measurement of the correlators $ \ev{ I(f)I(-2f) } \propto \chi_{-f}(f)$ and $ \ev{ I(f)I(f) } \propto \chi_{2f}(f)$ with $f=5.05$GHz, obtained with  $\tilde f\simeq -f$ or $\tilde f\simeq2f$.

\section{Detection scheme}

\begin{figure}[ht]
    \centering
    \includegraphics[width=6cm]{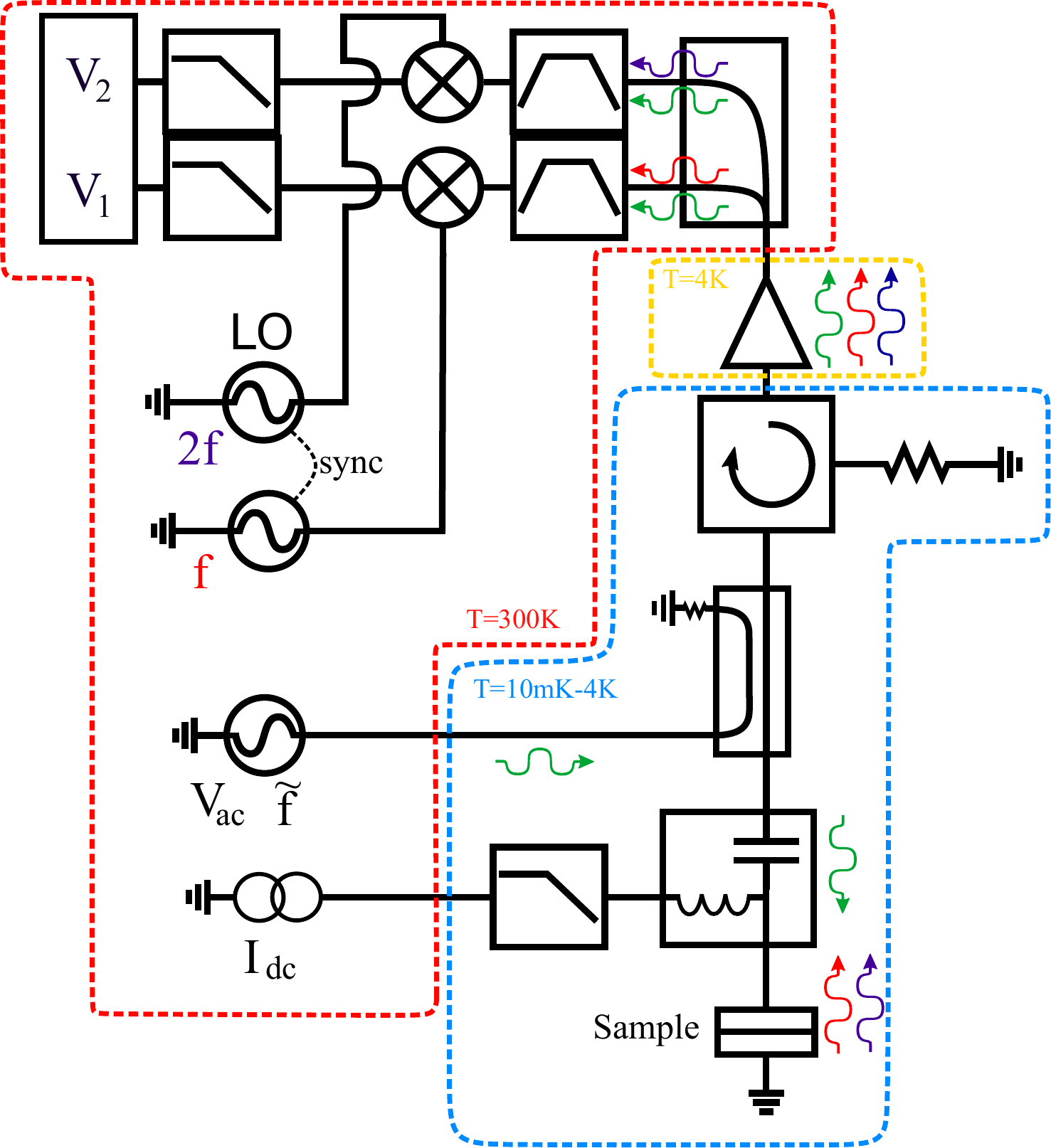}
    \caption{Experimental setup for the measurement of the noise susceptibility. Red (blue) arrows represent the current noise coming from the tunnel junction at frequency $f$ ($2f$). Green arrows represent the ac excitation at frequency $\tilde f$ sent towards the sample and reflected into the detection setup.}
    \label{exp_setup}
\end{figure}

The sample is a 93$\Omega$ tunnel junction fabricated using standard photo-lithography and metal deposition techniques \cite{Dolan1977}. It is placed on the 10mK stage of a dilution refrigerator. The detection setup is illustrated in Fig. \ref{exp_setup}. A bias tee separates the dc line used to apply a current bias ($I_{dc}$) in the sample, from the microwaves. A directional coupler allows to send the ac excitation towards the junction. The high frequency noise generated by the junction is amplified at 4K by a high electron mobility transistor cryogenic amplifier and further amplified at room temperature. A circulator is used to stop the noise coming from the amplifier and other high-temperature components to heat up the junction. The room temperature signal is split into two separate bands 4-6 GHz and 8-12 GHz by a diplexer followed by band pass filters. These signals are downconverted by mixing at frequencies $f=5.05$GHz and $2f$, with LOs provided by two phase-locked microwave sources. The low frequency fluctuations $V_1(t)$ obtained after mixing at $f$, and $V_2(t)$ obtained after mixing at $2f$ are digitized by synchronous 14bits 400MS/s A/D converters. From the stream of data we calculate the variances $\ev{V_1^2}\propto S_2(f)$ and $\ev{V_2^2}\propto S_2(2f)$ as well as the correlator $\ev{V_1^2(t)V_2(t)}$ which are averaged over time. 
The latter is proportional to $\mathrm{Re}\ev{V_{out}(f)^2V_{out}(-2f)}$ where $f$ is averaged over the bandwidth of the digitizer, i.e. 200MHz around $f$. $V_{out}$ is the voltage coming from the junction that is amplified and detected. We recall that since $V_{out}(t)$ is real, negative frequencies correspond to the complex conjugate: $V_{out}(-f)=V_{out}(f)^*$. The phase between the two oscillators is adjusted to maximize the signal. We checked that no signal is detected if the phase is rotated by 90 degrees (data not shown). This corresponds to $\mathrm{Im}\ev{V_{out}(f)^2V_{out}(-2f)}=0$, as expected. In the following we consider only the real part of all correlators and omit to write Re() in front of all correlators to simplify the notation.

Since the junction is not matched with the 50$\Omega$ microwave circuitry, it partially reflects the excitation. When the excitation frequency $\tilde f$ falls within one of the detection bands, it dominates the fluctuations within the corresponding band. Thus for $\tilde f\sim f$, the measured correlator is proportional to $V_{ac}(f)\ev{V_{out}(f)V_{out}(-2f)}\propto \abs{V_{ac}}^2\chi_{-f}(f)$ while for $\tilde f\sim 2f$ we measure $\abs{V_{ac}}^2\chi_{2f}(f)$ with $V_{ac}$ the ac voltage.

The non-linearities of the detection setup (mostly coming from mixers, amplifiers and the acquisition card) can introduce unwanted contributions in the measured signal. However these involve the detected power, i.e. the variance of current fluctuations, $\ev{ \abs{V_{out}(f)}^2 }$ and $ \ev{\abs{V_{out}(2f)}^2}$ \cite{Reulet2003}. Since these are even functions of the dc bias voltage $V$ while the signal of interest, the noise susceptibility, is odd in $V$, a simple anti-symmetrization of the total measured signal with respect to $V$ removes the effects of the non-linearities.

\section{Results}

\begin{figure}[ht]
    \centering
    \includegraphics[width=8cm]{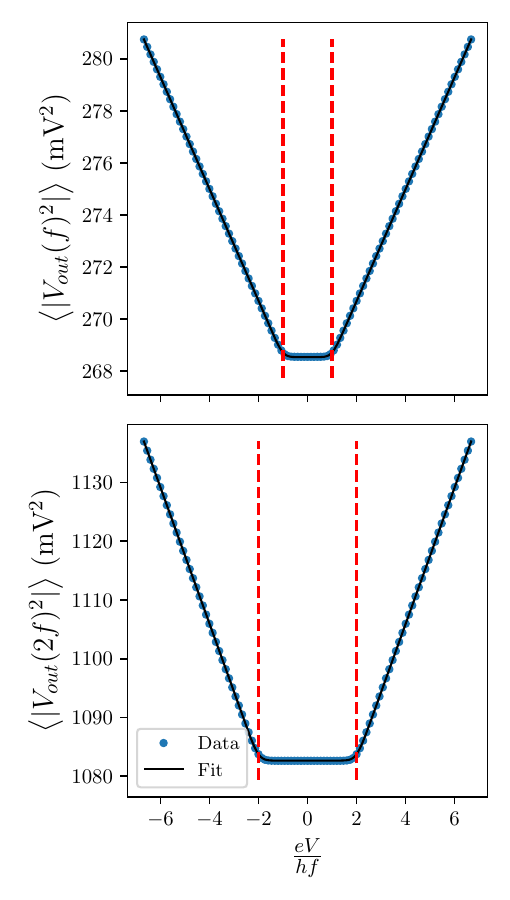}
    \caption{Measurement of $S_2$ at $f$ (top) and $2f$ (bottom) when no ac excitation is being sent to the sample. Blue symbols are data, black solid lines are theoretical fits with the electron temperature, the global gain $G_0$ and added noise $S_0$ as free parameters. The red vertical dashed lines highlight the characteristic plateau of the vacuum fluctuations, which width is given by the frequency of measurement; $\frac{eV}{hf} \pm 1$ for $S_2(f)$ and $\frac{eV}{hf} \pm 2$ for $S_2(2f)$.}
    \label{s2_mes_exc2f}
\end{figure}

We first consider the variance of current fluctuations in the absence of ac excitation. Fig. \ref{s2_mes_exc2f} shows the noise measured at frequency $f$ (top) and $2f$ (bottom) as a function of the rescaled dc voltage $eV/hf$. These are related to the noise spectral densities of current fluctuations generated by the junction by: $\langle|V_{out}(f)|^2\rangle=G(f)[S_2(f)+S_A(f)]$ (and a similar formula for $2f$). $G(f)$ represents the sensitivity of the measurement, which combines the attenuation along the microwave lines, reflection on the sample and gain of the detection setup. $S_A(f)$ is the noise added by the detection setup, mostly coming from the amplifier. As the shot noise of a tunnel junction can be used as a primary thermometer \cite{Spietz2003}, the electronic temperature $T$ is deduced by fitting the data of Fig. \ref{s2_mes_exc2f} (black lines) using\cite{BlanterButtiker}:

\begin{equation}
    S_2(f) = \frac{1}{2} \qty( S_2^0(f_+) + S_2^0(f_-))
    \label{S2_f}
\end{equation}
\noindent We find $T\simeq40$mK for both measurements. This temperature is low enough to be deep in the quantum regime, with $hf/k_BT\sim 12$ for $f=5.05$GHz. The plateaus observed for $e|V|<hf$ in $S_2(f)$ and $e|V|<2hf$ in $S_2(2f)$ correspond to vacuum fluctuations (between the red vertical dashed lines in Fig. \ref{s2_mes_exc2f}). In this bias regime no photon of energy $hf$ (resp. $2hf$) can be emitted if the energy of tunneling electrons $e|V|$ is smaller. At higher voltage, $S_2=eI$. The finite temperature corresponds to the rounding of the curves near $e|V|=hf$ (resp. $eV=2hf$).

\begin{figure}[ht]
    \centering
    \includegraphics[width=9cm]{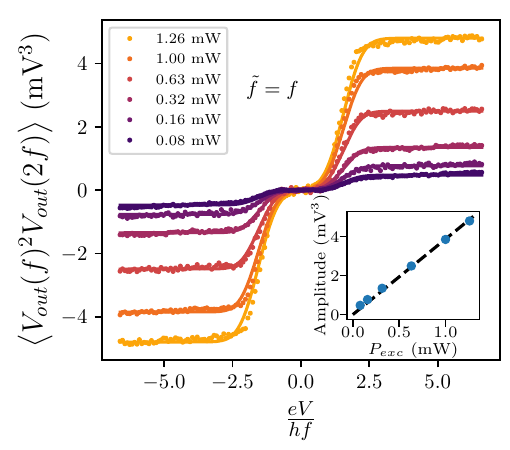}
    \caption{Measurement of $\chi_{-f}(f)$ vs. rescaled dc bias voltage, for various excitation powers. Symbols are data, solid lines are theoretical predictions with only the amplitude as a fitting parameter. Inset: amplitude of the signal at large voltage as a function of the ac power.}
    \label{chi_mes}
\end{figure}

\begin{figure}[ht]
    \centering
    \includegraphics[width=9cm]{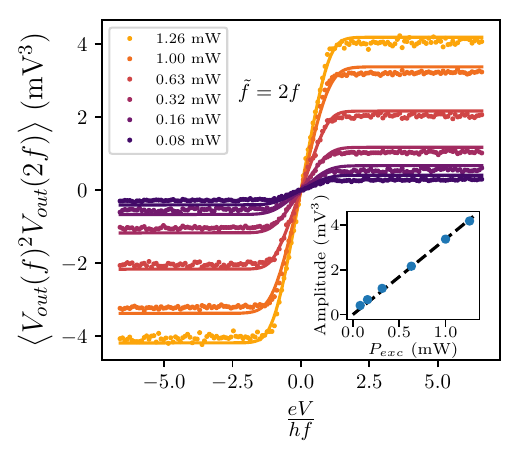}
    \caption{Measurement of $\chi_{2f}(f)$ vs. rescaled dc bias voltage, for various excitation powers. Symbols are data, solid lines are theoretical predictions with only the amplitude as a fitting parameter. Inset: amplitude of the signal at large voltage as a function of the ac power.}
    \label{chi_mes2}
\end{figure}

We now turn to the noise susceptibility. We show as symbols in Fig. \ref{chi_mes} the results for $\chi_{-f}(f)$ (excitation at $\tilde f \sim f$) and in Fig. \ref{chi_mes2} that for $\chi_{2f}(f)$ (excitation at $\tilde f \sim 2f$), both as a function of the dc bias voltage.

Using the theory described in \cite{Gabelli2008_theo}, with the known electron temperature and junction impedance, we can fit the data with one single parameter, the overall amplitude of the signal.  We observe that all the curves are very well fitted by the same theoretical curve up to an overall amplitude, see solid lines in both figures. We have checked, see insets in Figs. \ref{chi_mes} and \ref{chi_mes2}, that the amplitude of the measured signal scales linearly with the power of the ac excitation. Indeed, the response of the sample scales linearly with the ac voltage, and so does the excitation reflected by the sample. Thus the detected correlator is proportional to the ac excitation power. 

 At high bias $e|V| > 2hf$, both susceptibilities are constant, $\chi_{\tilde f}(f) = Ge$ independent of $f$ and $\tilde f$. In contrast when $e|V|<hf$ there is a remarkable difference in $\chi$ when we either send the excitation at $f$ or $2f$ : $\chi_{-f}(f)$ vanishes while $\chi_{2f}(f)$ grows linearly with dc bias.
 The two measured noise susceptibilities show also a clear deviation from the adiabatic regime where $\chi_{0}(f) = \frac{dS_2(f)}{dV}$, which is zero up to $V=hf/e$, then abruptly rises to $Ge$ on a voltage span of $\sim k_BT/e$. 
 In order to understand the bias dependence of the two susceptibilities, we take the standpoint of the photons generated by the junction \cite{Grimsmo2016} and invoke energy conservation. Exciting at $2f$ allows for one photon of energy $2hf$ to give birth to a pair of photons of energy $hf$, leading to $\ev{I(f)I(f)}\neq0$. Indeed, this correlator is the one involved in squeezing experiments \cite{Gasse2013,Bednorz2013}. This procedure, usually referred to as three wave mixing, occurs as soon as the nonlinear process that generates the photon pairs is not on a symmetry point, here $V=0$. As a consequence, $\chi_{2f}(f)\propto\ev{I(f)I(f)}$ grows as soon as $V\neq 0$. In contrast, the correlator $\ev{I(f)I(-2f)}$ involves photons at both $f$ and $2f$. Since here the pump is at $f$, an extra energy $hf$ is required to generate photons at frequency $2f$. This energy can be provided by single electrons only if $eV>hf$ (it could be provided by a second pump photon but we consider here the linear response in the ac excitation, i.e. neglect second order processes which may occur at higher pump power). As a consequence, $\chi_{-f}(f)$ is non-zero only when $ \abs{eV}>hf$. This attempt at understanding the behaviour of the noise susceptibility as a function of dc bias is however purely qualitative and would require a deeper analysis, in particular to understand how the processes of emission and absorption of photons are involved. 

In this experiment we have used an excitation of low, yet finite amplitude. At higher power the measured correlators are not simply proportional to the ac power \cite{Gabelli2008_theo,Gabelli2008_exp}. In the opposite limit, when no excitation is intentionally sent towards the sample, the latter still experiences the noise coming from the rest of the circuit and the one generated by the sample itself. These lead to environmental contributions to the third moment of voltage fluctuations which involve the noise susceptibilities \cite{Reulet2003, Beenakker2003}.

\section{Conclusion}
We have designed an experimental setup to measure a third order correlation between fluctuations at 5 and 10 GHz. This setup allowed us to perform the measurement of the noise susceptibility of a tunnel junction in the quantum regime, an important step towards the study of non-Gaussian quantum noise. The measurements are in very good agreement with theoretical predictions. 

\section*{Acknowledgments}
We would like to acknowledge Gabriel Laliberté and Christian Lupien for their technical help and time. This work was supported by the Canada Research Chair program,
the NSERC, the Canada First Research Excellence Fund, the FRQNT, and the Canada
Foundation for Innovation.

\bibliographystyle{IEEEtran}
\bibliography{IEEEabrv,biblio}

\end{document}